\documentclass[slac_one]{revtex4}
\usepackage{graphicx}
\usepackage{fancyhdr}
\usepackage{color}
\include{babarsym}
\def\Bztopizpiz  {\ensuremath{ B^0 \to \piz\piz}\xspace}

\def\Abar  {\kern 0.2em\overline{\kern -0.2em A}{}\xspace}

\begin{document}


\title{Charmless Two-Body and Quasi-Two-Body {\boldmath $B$}-decays at \babar} 

\author{Ingrid Ofte}
\affiliation{SLAC, 2575 Sand Hill Road, Menlo Park, CA 94025, USA \hspace{4cm} \\
Representing the \babar\ Collaboration}
\begin{abstract}
We present improved measurements of the branching fractions and \CP\
asymmetries in the two-body decays $B^0\to\pi^0\pi^0$, $B^0\to K^0\pi^0$ 
and $B^0\to K^+\pi^-$ as well as the quasi two-body 
$B^0 \to K_1(1270)^+ \pi^-$ and $B^0 \to K_1(1400)^+ \pi^-$ decays. 
These updated measurements are made using the complete set of \babar\ data 
taken at the Y(4S) resonance, collected between 1999 and 2007 at the PEP-II 
collider at SLAC. 
\end{abstract}

\maketitle

\thispagestyle{fancy}

\section{INTRODUCTION} 
$B$-mesons decaying into charmless hadronic final states are rare
events. The leading tree diagrams are CKM suppressed and loop
(penguin) diagrams typically contribute at comparable magnitude.  
Decay rates and \CP\ asymmetries may
deviate from Standard Model expectations if yet-unknown heavy
particles contribute in the loop.
The abundance of related decays allows for multiple independent
measurements of CKM matrix parameters and \CP\ violation. 

Direct \CP\ asymmetry occurs when the magnitude squared of a decay
amplitude differs from that of its \CP\ conjugate process, $|A(\bar B
\to \bar f)|^2 \neq |A(B \to f)|^2$.  Experimentally, we measure
$A_{\CP}$ in terms of different yield of one process compared to that
of the \CP\ conjugate process. We expect non-zero $A_{\CP}$ if two or
more amplitudes of comparable size contribute with different weak
($\phi$) and strong ($\delta$) phases, e.g. in the $B\to K\pi$ decay,
where we have similar-sized contributions from tree and penguin
amplitudes. The direct \CP\ asymmetry is given by
\( A_{\CP} =  2 \sin \phi \sin \delta / ( |T/P| + |P/T| + 2 \cos \phi \cos \delta ) \).            
Theoretically, the amplitudes ratio $|T/P|$ and phase $\delta$ are extremely
difficult to compute, as these quantities involve long-distance
effects. 

The hadronic uncertainties cancel to some extent in appropriately
constructed ratios. Sum rules~\cite{SumRules} based on isospin and
flavor SU$(3)$ symmetries relate decay rates and asymmetries in
different final states and can make precision tests possible in spite
of the hadronic uncertainties.

In these proceedings we present preliminary results from the analysis of
four charmless $B^0$ decay modes to two-body or quasi two-body final
states using data from the \babar~\cite{babar} detector. 
The analyses are described in detail in~\cite{telnov conf, K1pi}.

\section{ANALYSIS TECHNIQUE}
We use 467 million \FourS decays collected between 1999 and 2007,
which amounts to about 22\% more \BB\ pairs than used previously. 
Improvements in track reconstruction have
further increased the efficiency of our analyses and thereby improved
the statistical significance. For some of the decay modes, additional 
sensitivity is obtained from improved analysis techniques.

Charmless two-body and quasi two-body decays of $B$ mesons typically
have high reconstruction efficiency and large $\q \bar q$ backgrounds. To reduce this
background we make use of event-shape variables. At the PEP-II
collider, the $B$-mesons are produced almost at rest in the
center-of-mass (CM) frame, and their decay products are isotropically
distributed. Backgrounds from $e^+e^- \to q \bar q$ events, on the
other hand, are produced with larger momenta and have a more jet-like
event structure. To distinguish these from signal $B$ decays, we use
one or more of the following observables, all evaluated in the
CM frame: the sums $L_0\equiv\sum_i |{\bf p}^*_i|$ and
$L_2\equiv\sum_i |{\bf p}^*_i| \cos^2 \theta^*_i$, where ${\bf p}^*_i$
are the momenta and $\theta^*_i$ are the angles with respect to the
thrust axis~\cite{ref:thrust} of the \B candidate, of all tracks and
clusters not used to reconstruct the signal \B-meson candidate; $|\cos
\theta^*_{\scriptscriptstyle S}|$, where $\theta^*_{\scriptscriptstyle
  S}$ is the angle between the sphericity axes~\cite{Bjorken:1969wi}
of the \B candidate's decay products and that of the remaining tracks
and neutral clusters in the event; $|\cos\theta^*_B|$, where
$\theta^*_B$ is the angle between momentum vector of the signal $B$
and the beam axis; and $|\cos\theta^*_{\scriptstyle T}|$, where
$\theta^*_{\scriptstyle T}$ is the angle between the thrust axis of
the signal $B$-meson's daughters and the beam axis.

Correctly reconstructed $B$ decays are selected based on their
kinematic signatures exploiting the fact that each of the $B$ mesons
have half of the precisely-known beam energy. Most commonly we use the
energy substituted mass, $\mes = \sqrt{ (E^{\rm CM}_{\rm beam})^2 -
(p^{\rm CM}_{B})^2}$ and the energy difference, $\DeltaE = E_{B} -
\sqrt{s}/2$, where $\sqrt{s}$ is the total CM energy.  In the case of
$B^0\to K^0_S \pi^0$, we instead use a kinematic constraint. We define
$m_{\rm miss} = |\sqrt{s} - \hat{q}_B |$, where $\hat{q}_B$ is the
four-momentum of the reconstructed $B^0$ after a $B^0$ mass constraint
has been applied. Also used is $m_B$, the invariant mass of the signal $B$.

In the modes where the $B^0$ decays into a \CP\ eigenstate, $B^0_{\CP}$, we rely on 
a multivariate technique~\cite{ref:sin2betaPRL04} to determine the flavor of 
the other $B$, $B_{\rm tag}$.
Finally, we extract the signal yield and \CP asymmetries via an extended unbinned
maximum likelihood (ML) fit to the kinematic and event-shape variables.

\section{RESULTS}

\subsection{Branching fraction of {\boldmath $B^0 \to K^0 \pi^0$ }}
We reconstruct $K^0_S \to \pi^+\pi^-$ and $\pi^0\to \gamma\gamma$ and
measure the branching fraction via a ML fit to $m_{\rm miss}$,
$m_{B}$, $L_2/L_0$ and $\cos\theta_{B}^*$, as well as $\Delta t$,
the proper time difference between the decay of the signal
$B$ and that of the other $B$ in the event, to
enable extraction of \CP\ asymmetry~\cite{CP pi0K0S, jims proceedings}.  The fit extracts
a yield of $556 \pm 32$ signal events.  With a selection efficiency
$\varepsilon = (34.2\pm1.2)\%$, this translates into a branching
fraction measurement ${\cal B}(B^0 \to K^0 \piz ) = ( 10.1 \pm 0.6 \pm
0.4 ) \times 10^{-6}$, in agreement with, and superseding, \babar's
previous result~\cite{K0sPi0 PRD77 2008}.

Sum rules assuming flavor SU(3) symmetry~\cite{SumRules} can be used to 
make a prediction based on the other, more precisely measured, $K\pi$ 
final states: 
\(  2 \Gamma(B^0 \to K^0 \pi^0) + 
    2 \Gamma(B^+ \to K^+ \pi^0) = 
    \Gamma(B^+ \to K^0 \pi^+) +
    \Gamma(B^0 \to K^+ \pi^-)
\). Using this formula, we obtain the prediction 
\({\cal B}(B^0\to K^0 \pi^0)^{\rm Sum~rule} = (8.4\pm0.8)\times 10^{-6}\).
Our measurement is in reasonable agreement with this prediction within 
experimental uncertainties. 

\subsection{\boldmath Direct \CP\ asymmetry in $B^0 \to K^+\pi^-$}
\begin{figure}[htb]
  \begin{center}
  \setlength{\unitlength}{1.0cm}
  \begin{picture}(12,7)
    \put(0,0){\includegraphics[width=0.5\textwidth]{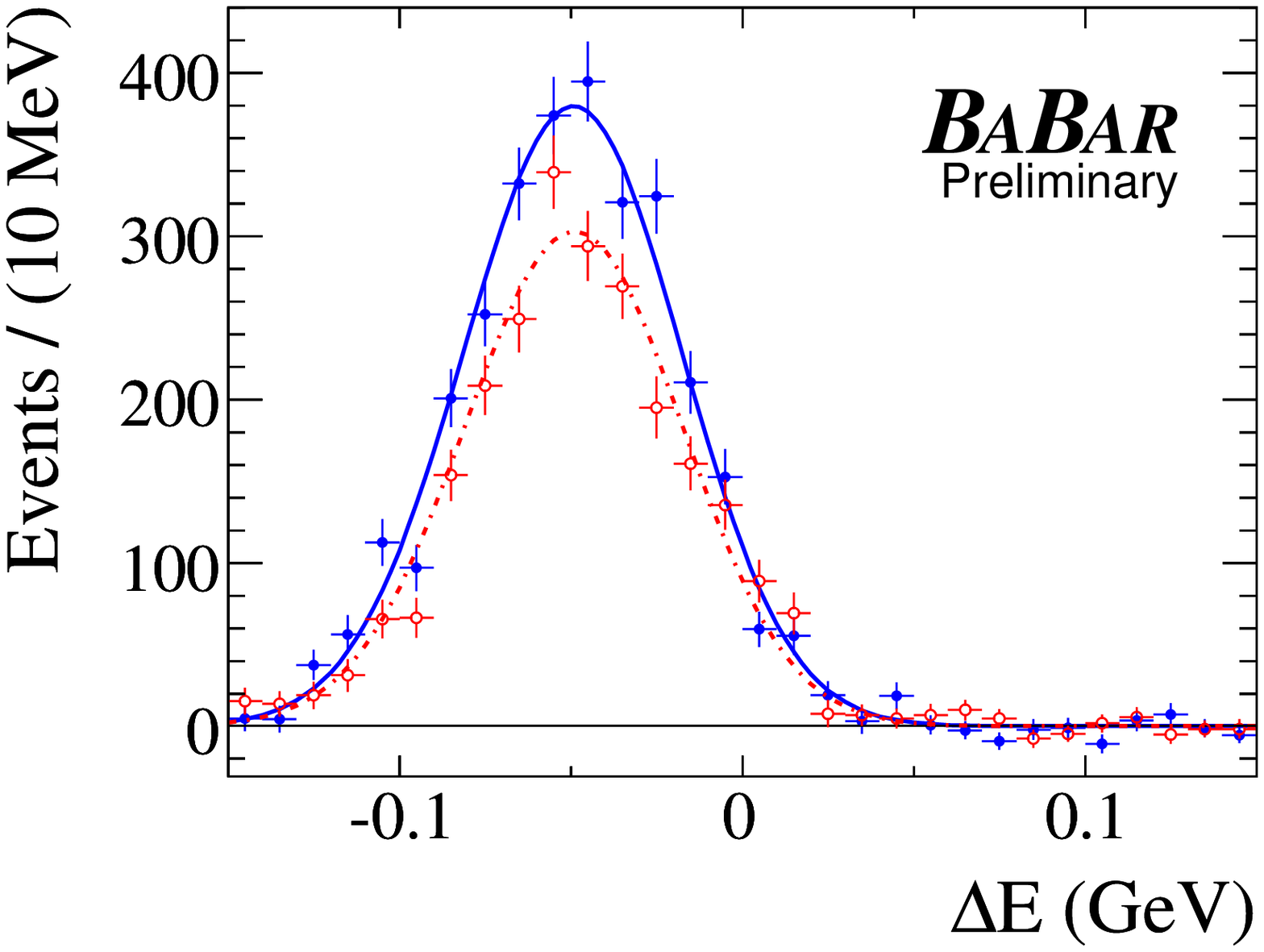}}
    \put(6.2,4.0){\color{blue} $B^0 \to K^+\pi^-$}
    \put(6.2,3.3){\color{red} $\overline{B^0} \to K^-\pi^+$}
    \put(5.5,4.1){\color{blue}\line(1,0){0.5}}
    \put(5.5,3.4){\color{red} \dashbox{0.06}(0.5,0.00){}}
  \end{picture}
\caption{$s{\cal P}$lot~\cite{sPlot} of \DeltaE for $K^{\pm}\pi^{\mp}$ events. 
The different decay rates for $B^0 \to K^+\pi^-$ and $\overline{B^0} \to K^-\pi^+$ are apparent. 
}\label{fig:Kpi}
\end{center}
\end{figure}

This decay is a self-tagging mode and the \CP\ asymmetry is apparent
from the different decay rates of $B^0 \to K^+\pi^-$ and $\overline{B^0} \to K^-\pi^+$.
We reconstruct the $B$-meson from two oppositely charged tracks that
are both assumed to be pions for the purpose of calculating
\DeltaE. The signal $\pi\pi$ and $K\pi$ yields are extracted from 
a ML fit to the kinematic variables \mes and \DeltaE, a 
${\cal F}$isher discriminant based on $L_2$ and $L_0$, and $\Delta t$ 
and $B$ flavor tagging in order to extract the time-dependent \CP\ 
parameters for $\pi^+\pi^-$ (see~\cite{Gritsan}). In addition,
particle-identification observables (the Cherenkov angle $\Theta_C$ in
the DIRC~\cite{dirc} and ionization-energy loss $dE/dx$ in the main
tracking detector) are used to separate $K$ tracks from $\pi$ tracks.
The $\Delta E$ distribution is offset from zero for $K\pi$ events,
seen in figure~\ref{fig:Kpi}, which further aids in distinguishing the
$\pi\pi$ and $K\pi$ events. The different rates for $B^0$ and
$\overline{B^0}$ is apparent from the figure and we measure
$A_{K^+\pi^-} = -0.107 \pm 0.016^{+0.006}_{-0.004}$ with $6.1\sigma$
significance, in agreement with and superseding our previous
result~\cite{PubKPi}.

\subsection{\boldmath Branching fraction and time-integrated \CP\ asymmetry in $B^0 \to \pi^0\pi^0$}

The decay $B^0 \to \pi^0\pi^0$ is useful for the extraction of the CKM
angle $\alpha$ from an isospin analysis of the $B \to \pi\pi$
system~\cite{Isospin}.  Combined with the other two $B\to \pi\pi$
modes, the rates and \CP\ asymmetries can determine $|\Delta \alpha| =
|\alpha - \alpha_{\rm eff}|$ with a four-fold ambiguity. This evaluation
has been presented in~\cite{Gritsan}.

We reconstruct the $B^0$ from a pair of $\pi^0$ candidates. $\pi^0\to
\gamma\gamma$ are formed from pairs of clusters in the EMC that are
isolated from any charged tracks. For this mode, we also use $\pi^0$
candidates from a single EMC cluster containing two adjacent photons
(a merged $\pi^0$), or one cluster and two tracks from a photon
conversion to an $e^+e^-$ pair inside the detectors. The yield and
time-integrated \CP\ asymmetry is obtained from a ML fit to the
kinematic variables \mes and \DeltaE, as well as the output of a
neural network based on event-shape variables. The use of a neural
network has improved the statistical sensitivity compared to the
previous result. The time-integrated \CP\ asymmetry is measured by
the $B$-flavor tagging algorithm.

We use an improved background model for the ML fit, where we allow the
background shape parameter to be linearly dependent on the neural
network output observable. With this, we obtain a better fit to the
data than in our previous measurement~\cite{pi0pi0_BaBar}.  We observe
$N_{\piz\piz} = 247\pm 29$, with an efficiency $\varepsilon =
(28.8\pm1.8)\%$. This results in the branching fraction measurement
${\cal B}(\Bztopizpiz)= ( 1.83 \pm 0.21 \pm 0.13 ) \times 10^{-6}$.
We also obtain the time-integrated \CP\ asymmetry measurement $A_{\CP}
= -C_{\piz\piz} = 0.43 \pm 0.26 \pm 0.05$.

\subsection{\boldmath Branching fraction of $B^0 \to K_1(1270)^+ \pi^-$ and $K_1(1400)^+ \pi^-$}

There has been interest recently in $B$-meson decays to an axial
vector and a pseudoscalar meson. Experiments have found relatively large
branching fractions and more measurements are needed to improve our 
understanding of these decays.  These $B^0$ decays are also of interest to
the measurement of the CKM angle $\alpha$. A measurement of
$\alpha_{\rm eff}$ can be obtained from $B^0 (\Bzb) \to
a_1(1260)^{\pm} \pi^{\mp}$~\cite{a1_1260_pi}.  Using SU(3)
flavor-symmetry~\cite{zupan}, theoretical bounds can be set on the
difference $\Delta \alpha = \alpha- \alpha_{\rm eff}$ by relating $B^0
(\Bzb) \to a_1^{\pm} \pi^{\mp}$ with $\Delta S = 1$ decays: $B \to a_1
K$ and $B \to K_{1A}\pi$, where $K_{1A}$ is a mixture of $K_1(1270)$
and $K_1(1400)$.

$K_1(1270)^+$ and $K_1(1400)^+$ are wide, overlapping axial
mesons. Both are reconstructed through their decays to $K^+\pi^+\pi^-$
final state. $B$ mesons are reconstructed from $B^0 \to K_1(1270)^+
\pi^-$ and $K_1(1400)^+ \pi^-$ in 454 million \BB pairs.  
This year's updated analysis uses an improved signal
model~\cite{K1model} to describe the production of $K_1(1270)$ and 
$K_1(1400)$, including interference effects. The
decay is described in terms of two real production parameters
$(\theta, \phi)$, related to the relative amplitude and phase
between $K_1(1270)^+$ and $K_1(1400)^+$.
We do a likelihood scan with respect to these two variables. For each 
likelihood point, we perform a ML fit. 
At the minimum $-ln {\cal L}$, we obtain a combined branching fraction 
${\cal B}(B^0\to K_1(1270)^+ \pi^- + K_1(1400)^+ \pi^-) = ( 31.0 \pm 2.7 \pm 6.9 ) \times 10^{-6}$.
We evaluate a significance of $5.1\sigma$. 
We also set limits on the ratio of the production constants for the 
$K_1(1270)^+$ and $K_1(1400)^+$ mesons in $B^0$ decays: 
$ 0.25 < \theta < 1.32$ and $-0.51 < \phi < 4.51$ at 95\% probability. 
This is the first attempt in $B$-decay data to measure the relative phase
between $K_1(1270)$ and $K_1(1400)$.

\section{CONCLUSION}

We present preliminary results on several charmless two-body and quasi two-body 
decays of neutral $B$-mesons based on the complete \babar\ data sample. 
We have three improved branching-fraction measurements:
${\cal B}(B^0 \to K^0 \piz ) = ( 10.1 \pm 0.6 \pm 0.4 ) \times 10^{-6}$, 
${\cal B}(\Bztopizpiz)= ( 1.83 \pm 0.21 \pm 0.13 ) \times 10^{-6}$ and 
${\cal B}(B^0\to K_1(1270)^+ \pi^- + K_1(1400)^+ \pi^-) = ( 31.0 \pm 2.7 \pm 6.9 ) \times 10^{-6}$. 
We have also two updated measurements of direct \CP\ asymmetry: 
$A_{\piz\piz} = -C_{\piz\piz} = 0.43 \pm 0.26 \pm 0.05$ and 
$A_{K^+\pi^-} = -0.107 \pm 0.016^{+0.006}_{-0.004}$. 
We have also made the first attempt to measure the relative production phase 
between $K_1(1270)$ and $K_1(1400)$ in the decays of $B$-mesons.

%
%
\begin{acknowledgments}
I would like to thank the organizers of ICHEP 2008 for
an interesting conference, and my \babar\ and PEP-II collaborators
for their contributions. This work was partially supported by U.S.\
Department of Energy and National Science Foundation.
\end{acknowledgments}

%
%


\begin{thebibliography}{99} 

\bibitem{SumRules}
H.~J.~Lipkin,
Phys.\ Lett.\  B {\bf 445}, 403 (1999)

M.~Gronau and J.~L.~Rosner,
Phys.\ Rev.\  D {\bf 59}, 113002 (1999)

\bibitem{babar}
  B.~Aubert {\em et al.} [\babar\ Collaboration],
  Nucl.\ Instrum.\ Methods Phys.\ Res. A {\bf 479}, 1 (2002).

\bibitem{telnov conf}
  B.~Aubert {\it et al.}  [\babar\ Collaboration],
  arXiv:0807.4226 [hep-ex].

\bibitem{K1pi}
  B.~Aubert {\it et al.}  [\babar\ Collaboration],
  arXiv:0807.4760 [hep-ex].

\bibitem{ref:thrust}
  S.~Brandt {\it et al.}, Phys.\ Lett.\ {\bf 12}, 57 (1964);
  E.~Farhi, Phys.\ Rev.\ Lett.\ {\bf 39}, 1587 (1977).

\bibitem{Bjorken:1969wi}
  J.~D.~Bjorken and S.~J.~Brodsky,
  \jprd{\bf 1}, 1416 (1970).

\bibitem{ref:sin2betaPRL04}
  B.~Aubert {\it et al.} [\babar\ Collaboration],
  \jprl{\bf 94}, 161803 (2005).

\bibitem{CP pi0K0S}
  B.~Aubert {\it et al.} [\babar\ Collaboration],
  arXiv:0809.1174 [hep-ex].

\bibitem{jims proceedings}
  J.~F.~Hirschauer, these proceedings.

\bibitem{K0sPi0 PRD77 2008}
  B.~Aubert {\it et al.}  [\babar\ Collaboration],
  Phys.\ Rev.\  D {\bf 77}, 012003 (2008)

\bibitem{Gritsan}
A.\ Gritsan, these proceedings. 

\bibitem{dirc}
I.~Adam {\it et al.}  [\babar-DIRC Collaboration],
  Nucl.\ Instrum.\ Meth.\  A {\bf 538}, 281 (2005).

\bibitem{PubKPi}
  B.~Aubert {\it et al.}  [\babar\ Collaboration],
  Phys.\ Rev.\ Lett.\  {\bf 99}, 021603 (2007).

\bibitem{sPlot}
M.~Pivk and F.~R.~Le Diberder,
  Nucl.\ Instrum.\ Methods Phys. Res., Sect. A {\bf 555}, 356 (2005).
%
\bibitem{Isospin}
M.~Gronau and D.~London, \jprl{65}, 3381 (1990).

\bibitem{pi0pi0_BaBar}
 B.~Aubert {\it et al.}  [\babar\ Collaboration],
  Phys.\ Rev.\  D {\bf 76}, 091102 (2007).

\bibitem{a1_1260_pi}
  B.~Aubert {\it et al.}  [\babar\ Collaboration],
  Phys.\ Rev.\ Lett.\  {\bf 98}, 181803 (2007)

\bibitem{zupan}
M.~Gronau and J.~Zupan, Phys.\ Rev.\ D 73, 057502 (2007)

\bibitem{K1model}
  C.~Daum {\it et al.}  [ACCMOR Collaboration],
  Nucl.\ Phys.\  B {\bf 187}, 1 (1981).

\end{thebibliography}
\end{document}